\documentclass[12pt]{article}
\begin{document}

\newtheorem{theorem}{Theorem}[section]
\newtheorem{lemma}[theorem]{Lemma}
\newtheorem{defn}{Definition}[section]
\renewcommand{\theequation}{\arabic{section}.\arabic{equation}}
\newcommand{\tightlist}[1]{\begin{list}{$\bullet$}
                        {\usecounter{enumi}\setlength{\parsep}{1pt}}
                          #1\end{list}}
\newcommand{\innerlist}[1]{\begin{list}{$\circ$}
                        {\usecounter{enumii}\setlength{\parsep}{1pt}
                          \setlength{\leftmargin}{8truept}}
                          #1\end{list}}
\newcommand{\numberlist}[1]{\begin{list}{(\roman{enumi})}
                        {\usecounter{enumi}\setlength{\parsep}{1pt}
                          \setlength{\leftmargin}{16truept}}
                          #1\end{list}}
\newcommand{\widelist}[1]{\vspace{1cm}{\noindent}
                         \begin{list}{$\bullet$}
                        {\usecounter{enumi}\setlength{\leftmargin}{8truept}}
                         #1\end{list}}
\def\sqr#1#2{{\vcenter{\vbox{\hrule height.#2pt
        \hbox{\vrule width.#2pt height#1pt \kern#1pt
           \vrule width.#2pt}
        \hrule height.#2pt}}}}
\def\quod{\hfill$\sqr84$\break}
\def\qed{\begin{flushright}$\sqr84$\end{flushright}}

\newcommand{\remarks}[1]{\vspace{0.2cm}{\noindent\large\textbf{Remarks:}}
                         \begin{list}{\arabic{enumi}.}
                        {\usecounter{enumi}\setlength{\leftmargin}{8truept}}
                         #1\end{list}}
\newcommand{\beq}{\begin{equation}}
\newcommand{\eeq}{\end{equation}}
\newcommand{\be}{\begin{eqnarray}}
\newcommand{\ee}{\end{eqnarray}}
\newcommand{\bes}{\begin{eqnarray*}}
\newcommand{\ees}{\end{eqnarray*}}

\newcommand{\nbf}[1]{\noindent\textbf{#1}}
\newcommand{\Pa}{Painlev\'e}
\newcommand{\Complex}{\textrm{\kern.24em \vrule
     width.02em height1.2ex
     depth-.05ex\kern-.26em C}}
\newcommand{\Real}{\textrm{I\kern-.20em R}}
\newcommand{\C}{{\mathcal C}}
\newcommand{\N}{{\mathcal N}}
\newcommand{\Z}{\widetilde{\mathcal Z}}
\newcommand{\Sd}{\widetilde{\mathcal S}}

\newcommand{\bi}{\begin{itemize}}
\newcommand{\ei}{\end{itemize}}
\newcommand{\bd}{\begin{description}}
\newcommand{\ed}{\end{description}}
\newcommand{\ic}{\item[$\circ$]}

\newcommand\bPhi{\overline{\Phi}}
\newcommand\hPhi{\hat{\Phi}}
\newcommand\hu{\hat{u}}
\newcommand\bz{\overline{z}}
\newcommand\om{\omega}
\newcommand\de{\widetilde{\delta}}
\newcommand\bo{\overline{\omega}}
\newcommand\tilphi{\widetilde{\phi}}
\newcommand\tilpsi{\widetilde{\psi}}
\newcommand\tildo{\widetilde{\omega}}
\newcommand\bartildo{\overline{\widetilde{\omega}}}
\newcommand\primo{{\omega}'}
\newcommand\barprimo{\overline{{\omega}'}}
\newcommand\tildJ{\widetilde{J}}
\newcommand\tw{\widetilde{w}}

\newcommand\YE{{\om}_1{\bo}_2-{\bo}_1{\om}_2}
\newcommand\Y{{\tildo}_1{\bo}_2-{\bo}_1{\tildo}_2}
\newcommand\K{{\tildo}_1{\om}_2-{\om}_1{\tildo}_2}

\newcommand{\PI}{$\textrm{P}_{
       \textrm{\footnotesize I}}$}
\newcommand{\PII}{$\textrm{P}_{
       \textrm{\footnotesize II}}$}
\newcommand{\PIII}{$\textrm{P}_{
       \textrm{III}}$}
\newcommand{\PIV}{$\textrm{P}_{
       \textrm{IV}}$}
\newcommand{\PV}{$\textrm{P}_{
       \textrm{V}}$}
\newcommand{\PVI}{$\textrm{P}_{
       \textrm{VI}}$}

\title{The Second Painlev\'e Equation
in the Large-Parameter Limit I: Local Asymptotic Analysis}
\author{Nalini Joshi\thanks{Research supported
by the Australian Research Council.}\\
        {\it Department of Pure Mathematics}\\
        {\it University of Adelaide}\\
        {\it Adelaide Australia 5005}\\
  	 {\tt njoshi@maths.adelaide.edu.au}}
\maketitle
\begin{center}
Short Title: Large-parameter Local Asymptotics of \PII
\end{center}
\begin{center}
{\bf Abstract}
\end{center}
In this paper, we find all possible asymptotic behaviours of the solutions
of the second \Pa\ equation
$y''=2y^3+xy +\alpha$ as the parameter
$\alpha\to\infty$ in the local region $x\ll\alpha^{2/3}$.
We prove that these are asymptotic behaviours by finding explicit
error bounds. Moreover, we show that they are connected and complete
in the sense that they correspond to
all possible values of initial data given at a point in the
local region.

\section[1]{Introduction}
The asymptotic analysis of the six nonlinear second-order
ODEs (ordinary differential equations) known as the \Pa\ equations
\cite{ince} has been
a subject of both classical \cite{boutroux:I,boutroux:II} and modern
\cite{kitaev:survey,njmdk:conn1} interest. More recent interest was
revived by their importance
in physical applications, particularly
in statistical mechanics (see \cite{mccoy:92}), and by their close
relationship to soliton equations (see \cite{as:siam,ac:cup}).

The solutions of the \Pa\ equations
are, in general, highly transcendental functions known
as the \Pa\ transcendents. An advantage of asymptotic
description is that they give explicit behaviours of such transcendental
solutions in terms of known classical functions.
Such descriptions also give insight
into the asymptotic behaviours of the solutions of soliton equations
\cite{as:kdv,sa:p2}.

In this paper, we focus on the asymptotic analysis of
the second \Pa\ equation (or \PII )
\beq
y''=2y^3+xy + \alpha
\label{P2}
\eeq
in the limit as $\alpha\to\infty$. \PII\ is well known
to be a symmetry reduction of the modified Korteweg-deVries
equation (and other soliton equations,
see e.g. \cite{ac:cup}). In such a reduction,
the parameter $\alpha$
arises as a constant of integration. Therefore, a large
value of $\alpha$ is related to large values of initial data for
the soliton equation.

Although there have been many studies
\cite{boutroux:I,boutroux:II,sa:p2,njmdk:second,
cm:92,njmdk:conn1,ifk:94,kitaev:survey,dz:p2,bclm:unif}
of the \Pa\ equations in the limit as their independent variable
approaches a singularity of the equation, the asymptotic limit as
a \textit{parameter} approaches infinity has not been widely studied,
with two exceptions.
Kawai, Takei, \textit{et al} \cite{kt:I,kt:II,kt:0}
have studied the \Pa\ equations extensively
in such limits through their associated linear equations, called
isomonodromy problems \cite{okamoto:86,fuchs:annalen}.
Kitaev \cite{kitaev:92} has studied the double asymptotic limit of
the second \Pa\ equation as both $x\to\infty$ and $\alpha\to\infty$,
again through an isomonodromy problem. However, both
studies appear to be restricted to formal
special solutions.

In this paper, we prove that in a region bounded by $x\ll\alpha^{2/3}$
the local asymptotic behaviour of the general solution as $\alpha\to\infty$
is given by elliptic functions. We also show that there exist special solutions
whose local asymptotic behaviours are given by degenerate elliptic functions,
i.e. singly periodic functions. These results
are analogous to
the results found by Boutroux \cite{boutroux:I,boutroux:II} for the
\Pa\ equations in the limit as $x\to\infty$.

However,
the domains of validity are quite different from those that
are known for the limit $x\to\infty$.
In our case, these domains are full disks (punctured at poles
and stationary points) of large radius, unlike the case $x\to\infty$ in which
such domains are either bounded local patches
or sectors of angular width strictly less than $2\pi$.

Our methods are direct and do not use the relation of
\PII\ to an isomonodromy problem.
Part of our motivation is to
show that rigorous general asymptotic results can be obtained
through natural direct methods without going through associated
linear problems.

However, we do rely on a characteristic property, i.e.
the {\em \Pa\ property}, of the
\Pa\ equations to obtain sharp error bounds. This is the property
that all movable singularities of all solutions of each \Pa\
equation are poles. In particular, we use the direct proof
of the \Pa\ property of the \Pa\ equations given in \cite{njmdk:direct},
to define local paths of integration and to estimate $w$ on such paths.
Error bounds can be obtained
without assuming this property (see Boutroux\cite{boutroux:I}).
However, its usage allows us to obtain improved error
estimates.

The \Pa\ equations also possess many other deep properties.  For example,
their solutions can be written in terms of entire functions
\cite{pp:bull}, they possess exact special solutions in terms of
classical special functions and transformations relating solutions to
other solutions (see \cite{ac:cup} for references).
Asymptotics is a natural part of their study as nonlinear special functions.
\subsection{Asymptotic Regions}
The study
of \PII\ for $\alpha\to\infty$ necessarily decomposes into three
parts corresponding to three regions. First, consider a
transformation of \PII\ given by a
dominant-balance analysis as $\alpha\to\infty$, namely
\be
y(x)&=&\alpha^{1/3}w(z)\nonumber\\
z&=&\alpha^{1/3}x
\ee
which transforms \PII\ to
\beq
w''=2w^3+1+{z\over\alpha}w,
\label{aP2}
\eeq
where the primes denote $z$-derivatives. The three asymptotic regions
are given by
$z\ll\alpha$, $z\sim\alpha$, and $\alpha\ll z$. In this paper, we
restrict ourselves to obtaining rigorous
local asymptotic results in the first region.  Note that this corresponds
to $x\ll\alpha^{2/3}$.

These local results can be extended into the second region by using a
nonlinear multiple-scales approach developed by Joshi and Kruskal
\cite{njmdk:second,njmdk:conn1} for the first and second \Pa\ equations.
The results give insight into a
well known asymptotic limit of \PII . Consider \cite{ince}
the following transformation of \PII
\bes
x&\mapsto&\epsilon^2 X-{6\over \epsilon^{10}}\\
y(x)&=&\epsilon Y(X) +{1\over \epsilon^5}\\
\alpha&=&{4\over\epsilon^{15}}
\ees
In the limit as $\epsilon\to0$, this leads to
the first \Pa\ equation (\PI )
\beq
Y''=6Y^2+X.
\label{P1}
\eeq
It will be shown in a subsequent paper that this limit
occurs at isolated
points in the second region.

The method of Joshi and Kruskal
was developed for the limit $x\to\infty$. The usual asymptotic studies
of the \Pa\ equations in the limit $x\to\infty$
apply here in the third region given by $\alpha^{2/3}\ll x$.
To see this, consider the Boutroux transformation
of \PII\ that makes the maximal dominant balance as $x\to\infty$ explicit,
namely
\be
y(x)&=&\sqrt xu(t)\nonumber\\
t&=&{2\over 3}x^{3/2}
\ee
under which \PII\ becomes
\beq
u_{tt}=2u^3+u+{a-u_t\over t}+{u\over 9t^2},
\label{bP2}
\eeq
where $a=2\alpha/3$. (See \cite{njmdk:conn1} for a deduction of Boutroux
transformations based on maximal dominant balances.) The asymptotic
analysis of this equation is usually carried out for $t\to\infty$.
It remains valid so long as $a/t\ll 1$, i.e. $\alpha^{2/3}\ll x$.

\subsection{Main Results}
In section 2, we study the generic solutions of \PII .
These are solutions defined
by a large open set of initial data given at a point $z_0$ in the local
region. Such solutions are shown to
have asymptotic behaviours given by elliptic functions as $\alpha\to\infty$.
In section 3, we study limits of the generic solutions and their
asymptotic behaviours. The union of the results of Sections 2 and 3
cover all solutions defined by all possible initial data at $z_0$.

Moreover, the domain of validity of the generic locally valid
elliptic-function-behaviour consists of the entire local
domain $z\ll\alpha$ (punctured at poles and stationary points of the
solution). Similarly, its degenerate limits are also
valid in large subdomains of this region.

The solution in the nearly degenerate case is described by
an almost-singly-periodic behaviour, exhibiting
widely spaced lines of poles. There are further limits which exhibit
only one line of poles or no such line within the domain of validity.
We also consider the
large energy limit and show that the resultant behaviours are
still given by (scaled) elliptic functions in large subdomains.

Our results are given for both the solution and its first-derivative.
By using \PII , these may be extended to yield
the asymptotic behaviours of any higher-order derivative of $w(z)$.

In summary, we find all possible asymptotic behaviours of true
solutions in the limit as $\alpha\to\infty$ in large domains
lying in the region $z\ll \alpha$. We prove that these
are asymptotic behaviours by finding explicit error bounds.
We show that these
behaviours are connected through limiting values taken
by an energy-like parameter and are complete in the sense that
they correspond to all possible values of initial data
given at a point $z_0$ with $z_0\ll \alpha$.
\subsection{Notation}
It should be noted that we use two symbols for
an asymptotic balance between two functions $f(s)$ and $g(s)$ as $s\to\infty$
(along some path in the complex $s$-plane). First, if the limit of
$f/g$ is unity
then we write  $f\approx g$.
Second, if a nonzero limit exists but differs possibly from unity, we write
$f\sim g$.
Our remaining usage of asymptotic notation is standard and definitions may be
found, for example, in \cite{bo:asympt}.

\section{Generic Local Asymptotic Estimates}
In this section, we prove that
in the region where $z\ll\alpha$, the general solution of \PII\ is asymptotic
to
Jacobi elliptic functions. In particular, we estimate the size of the
asymptotic
correction term. Moreover, we show that the spacing
between zeroes of the general
solution of \PII\ is closely approximated by a period of
the locally valid elliptic function.
It should be noted that we obtain an asymptotic result
for both $w'(z)$ and $w(z)$.
Note that we assume $|\alpha|>1$ throughout.

First, we describe
the local domain for our asymptotic analysis.
\begin{defn}
Suppose $\epsilon\not=0$, $z_0$ are given complex numbers and $B$ is a positive
real number. They are called {\em appropriate} if
\bes
|\epsilon|&<&1, {|\ln\epsilon|\over|\alpha|}<|\epsilon|\\
\left|{z_0\over\alpha}\right|&<&|\epsilon|,
{B\over|\alpha|}< 2|\epsilon|\\
{(6+2\pi B)2\pi B}&<& |\epsilon|^{-5/7}|\ln\epsilon|^{-2}
\ees
For an appropriate set $\{\epsilon, z_0, B\}$
\[
{\mathcal Z}:=\{ z\Bigm| |z-z_0|\le B\},
\]
is called the {\em local $z$-domain}.
\label{app}
\end{defn}
\remarks{
\item The last inequality above is imposed to ensure that the error estimates
obtained in Theorem 2.1 below is small. See further remarks following that
theorem.
\item Note that one choice of appropriate $\{\epsilon, z_0, B\}$ is given by
$\epsilon=1/\sqrt{\alpha}$ with $z_0\sim {\alpha}^{1/4}$,
$B\sim {\alpha}^{1/8}$.
\item The above remark shows that $z_0$ and $B$ do not have to be small.
In particular, the local domain may be quite large (but not unboundedly large)
even if $z_0$ is
finite.}

Standard theorems \cite{hille:complex} show that for given bounded initial data
$w(z_0)$, $w'(z_0)$ at a
finite point $z_0$ there exists a unique locally
analytic solution $w(z)$ of \PII .
It is well known that $w(z)$ can be meromorphically
extended \cite{njmdk:direct}
to any bounded domain.
In particular, $w(z)$ is meromorphic in the local domain $\mathcal Z$.

We will work with an integrated form of \PII .
Multiply Eqn(\ref{aP2}) by $w'$ and integrate as though the small term
$z\,w/\alpha$ were not there. Then, we get
\beq
w'^2=w^4+2w+2E+2\int_{z_0}^z\,{zww'\over\alpha}\,dz,
\label{firstP2}
\eeq
where $E$ is a constant of integration. To leading-order, this equation
is solved by Jacobian elliptic
functions.

\subsection{Elliptic Functions}
To see that elliptic functions arise, and to fix the notation, let
\[P(w):=w^4+2w+2E.\]
In general, $P(w)$ has four distinct roots, which we will denote by $w_i$,
$i=1, \ldots, 4$, in the complex plane. The leading order equation
\beq
w'^2=w^4+2w+2E,
\eeq
can be integrated again to give
\beq
\int_\eta^w{dv\over \sqrt{P(v)}}=z-z_0,
\label{elliptic_integral}
\eeq
where $\eta=w(z_0)$. The path of integration here lies on
the Riemann surface defined by $\sqrt{P(v)}$. (This consists
of two copies of the complex plane punctured at the roots of $P(v)$ and glued
along lines  connecting two pairs of roots.)

There exist two linearly
independent closed contours $C_j$, $j=1,2$, on this surface each
enclosing two roots of $P(v)$ in the $v$-plane. Paths connecting
$\eta$ to the same end-point $w$ are deformable to each other modulo
$C_j$. That is, deformations lead to Eqn (\ref{elliptic_integral})
where $w$ remains unchanged while the right side
changes by integer multiples of
\beq
\om_j =\oint_{C_j}{dv\over \sqrt{P(v)}}.
\label{om}
\eeq
Further properties of the
elliptic integrals $\om_j$,
including differential equations they satisfy as functions
of $E$, are given in Appendix A.

This shows that the solution $w(z)$ (the inverse of
the function $z(w)$ defined by Eqn(\ref{elliptic_integral}))
has two periods given by $\om_j$. Moreover, it is straightforward
to show that $w(z)$ is meromorphic. That is, it is an elliptic function.

One of the two periods of $w$ becomes infinite (while the other remains
finite) at special values of
$E$. These are the values for which two roots of $P(w)$ coincide.
To find these, consider
\beq
P'(w) = 4w^3+2 =0\ \Rightarrow\ w= d_k:=\bigl(-1/2\bigr)^{1/3}, k=1, 2, 3.
\label{dblroots}
\eeq
If such points are also zeroes of $P(w)$ then we must have
\beq
E= D_k:=-{3\over 4}\bigl(-1/2\bigr)^{1/3}= \left({27\over 128}\right)^{1/3},
k=1, 2, 3.\label{Dk}
\eeq
For each such value of $E$, $P(w)$ has a double root. Note, however, that
$P(w)$
cannot have a triple root, because $P''(w)=12w^2$ cannot vanish simultaneously
with $P'(w)$.

Suppose $E=D_k$, for some $k = 1, 2$, or $3$. Let $\rho$ and $\sigma$ be the
roots of $P(w)$ distinct from $d_k$. Then rewriting
\[
P(w)=(w-d_k)^2(w-\rho) (w-\sigma)
\]
in expanded form shows that
\[
\rho\sigma=-{3\over 2d_k},\quad \rho+\sigma={1\over d_k^2}.
\]
Solving these equations, we get
\[
\rho, \sigma = {1\over 2d_k^2}\left\{1\pm i\sqrt{2}\right\}.
\]
Note that these cannot equal $d_k$ because $P(w)$ cannot have a triple root.
That is, there exists $c>0$, such that $|\rho-d_k|>c$, $|\sigma-d_k|>c$.

Now
suppose $E=D_k+\delta^2$, for $\delta\ll 1$. Then setting
$P(w)=0$ and using the fact that $d_k-\rho$, $d_k-\sigma$ are bounded below, it
is straightforward (see \cite{nj:corsica} for detailed proofs of
similar results)
to show that the two nearly coalescing roots $w_j$ of $P(w)$ satisfy
\[w_j=d_k+O(\delta),\]
and that the other two roots are a finite distance away from $d_k$.

Note also that if $E$ is large, say $|E|=\log\delta$, then the periods
become small. To see this, consider the roots of $P(w)$. The roots now have
modulus at least equal to $O(|E|^{1/4})$ and so
\[
\om_j=O(|E|^{-3/4}), j= 1, 2.
\]
(Such asymptotic results for the complete elliptic integrals can be obtained
straightforwardly from the differential equations derived in Appendix A.)
Hence the poles of the elliptic function crowd together and the function itself
becomes large even at points of holomorphy.

\subsection{Generic Solutions}
Our main result holds for solutions that are described by a large
open and dense set of initial values that we call {\em generic}.
\begin{defn}
Suppose $\epsilon$ is an appropriate given number. Parameters
$\eta$, $\eta'$ are called {\em generic} if for
\[2E= \eta'^2 - \eta^4 - 2\eta,\]
they satisfy
\[
|\epsilon|^{1/7}<|\eta'|<|\ln\epsilon|^2,\quad
|\eta|<|\ln\epsilon|
\]
\[
2|\epsilon|^{2/7}<|E-D_k|
\]
\end{defn}
These supply the initial values for the solutions and ensure that their
formal elliptic-function-type behaviours are not too close to degenerate ones.
Initial values that violate this definition are studied in section 3.
\begin{defn}
Suppose $\epsilon$, $z_0$ are appropriate numbers. Let $w(z)$ be
a solution of \PII\ defined by generic initial values $\eta$, $\eta'$
at $z_0$, i.e. $w(z_0)=\eta$, $w'(z_0)=\eta'$. Then $w(z)$ is called
a {\em generic} solution.
\end{defn}
To get valid asymptotic results, we need to confine such a solution $w(z)$
to domains that avoid its poles and stationary points.
\begin{defn}
Given an appropriate $\epsilon$, define the $w$-domain
by
\[
\mathcal W := \{ w\Bigm| |w|<|\ln\epsilon|, |P(w)|>|\epsilon|^{1/7}\}
\]
and $z$-domain by
\[
\Z := \{ z \in{\mathcal Z}\Bigm| w(z)\in \mathcal W\}.
\]
$\epsilon$ is called {\em small enough} if $\Z$ is connected.
\end{defn}
The direct proof \cite{njmdk:direct} of the \Pa\ property of \PII\ shows that
there exists an open disk of values $\epsilon$ for which
$\Z$ is connected. (Explicit lower bounds for $A:=1/|\epsilon|$,
may be found on p. 197 of \cite{njmdk:direct}.)
Moreover, any two points in $\Z$ are connected by paths $\gamma$, of
bounded length $2\pi B$, such that their images $\Gamma$ under $w$ lie in
$\mathcal W$ with length $<2\pi |\ln\epsilon|$.

Our major result is
\begin{theorem}
Given an appropriate set $\{\epsilon, z_0, B\}$ where
$\epsilon$ is small enough,
and generic initial values
$\eta$, $\eta'$,
the generic solution $w(z)$ satisfies
\bes
\left|w'^2 -P(w)\right|&\le& {\kappa\over 2}|\epsilon|\,|\ln\epsilon|^2\\
\left|\int_\eta^w\,{dv\over\sqrt{P(v)}
   }-(z-z_0)\right| &\le& 2\pi \kappa B |\epsilon|^{6/7}|\ln\epsilon|^2
\ees
for all $z\in\Z$, where $\kappa :=6+2\pi B$.
Moreover, if $z_0$ and $z_{10}$ are two
successive points in $\Z$ where $w=\eta$, then for
$j=1$ or $2$,
\[
\left|(z_{10}-z_0)-\omega_j\right|\le 2\pi \kappa B
|\epsilon|^{6/7}|\ln\epsilon|^2.
\]
\label{genthm}
\end{theorem}
\nbf{Proof:}
Equation(\ref{firstP2}) can be written
\[
\bigl(w'(z)\bigr)^2 = P\bigl(w(z)\bigr) + 2S
\]
where
\beq S:=\int_{\gamma} {z\over\alpha} w w'\,dz\label{s}\eeq
where $\gamma$ is a path connecting $z_0$ to $z$ in $\Z$. The
assumptions give
\[
\left|{z\over\alpha}\right|\le
{B\over|\alpha|}+\left|{z_0\over\alpha}\right| < 3|\epsilon|.
\]
Integration by parts shows that
\[S = \left[{zw^2\over 2\alpha}\right]_{(z_0,\eta)}^{(z, w)} - {1\over 2\alpha}
\int w^2 dz.\]
Hence we get
\bes
2|S|&\le& 6|\epsilon|\,|\ln\epsilon|^2
+ |\epsilon|\cdot 2\pi B|\ln\epsilon |^2\\
&=& \kappa |\epsilon|\,|\ln\epsilon|^2
\ees
as desired.
Now integrate Eqn(\ref{firstP2}) once more by taking the square root and
dividing  by $\sqrt{P(w)}$. We get
\beq
\int_\eta^w\,{dv\over\sqrt{P(v)}}=\int_{\gamma}\,dz
                  \left\{1+{2S\over P(w)}\right\}^{1/2}.
\label{secondP2}
\eeq
Note that we fix a choice of branch of the square root here, so that as
$\epsilon\to 0$ we have the positive real branch. From Eqn(\ref{secondP2})
we get
\beq
\int_\eta^u\,{dv\over\sqrt{P(v)}}-(z-z_0)=
   \int_{z_0}^z\,dz\left\{\left(1+{2S\over P(w)}\right)^{1/2}-1\right\}.
\label{thirdP2}
\eeq

Note that
\beq
\left|{2S\over P(w)}\right|\le \kappa |\epsilon|^{6/7}|\ln\epsilon|^2.
\eeq
Let
\[
Q:= \left(1+{2S\over
P(w)}\right)^{1/2}+1.
\]
Then we have
\[
Q^2=2Q+{2S\over P(w)}\Rightarrow\ Q=2+{2S\over Q P(w)}.
\]
The choice of branch ensures that $Q$ is lower
bounded by $|Q|\ge 1$.
Therefore, it follows that
\[
|Q-2|\le\left|{2S\over P(u)}\right|\quad\Rightarrow\ \left|\sqrt{1+{2S\over
P(u)}}-1\right|\le \kappa |\epsilon|^{6/7}|\ln\epsilon|^2.
\]
Integration then gives
the desired result.  The separation
$z_{10}-z_0$ is based on the same result with the path of integration $\Gamma$
being a closed contour enclosing two roots of $P(u)$. \quod

\remarks{
\item The crucial element of this proof lies in the lower bound
for $P(w)$. The microscopic details of this lower bound can be changed without
affecting the macroscopic form of the asymptotic result.
For example, suppose the conditions on $\eta$, $\eta'$ are changed to
allow $E-D_k=\delta^2$ where $|\delta|\sim|\epsilon|^{p/7}$, $p < 7$.
Then the same result as in the theorem holds for $w'^2$, and the second (and
third) results of the theorem hold with the
order of the error estimate changed to
\[|\epsilon|^{1-p/7}|\ln\epsilon|^2.\]
\item It should be noted, however, that if $\delta$ is small enough (but
nonzero), then
the incomplete elliptic integral on the left of the first result is
more accurately written as the inverse of a hyperbolic
function. See the next section for details.
\item The third result of the theorem holds for any two successive points
(in a direction given by $\omega_j$ for a chosen $j$) at which $w$
takes the same value. By taking $\epsilon$ to be smaller, such points can
be taken arbitrarily close to poles of $w$. Hence the result also holds for
the spacing between two poles.
\item Notice that the incomplete elliptic integral
\[ \int_\eta^w\,{dv\over\sqrt{P(v)}
   }\]
is still convergent if the generic initial value $\eta$ happens
to be arbitrarily close to
a zero of $P(w)$. However, if it is so close to such a zero that the
path intrudes into a \lq\lq hole\rq\rq\ of $\Z$,
then we need a slightly modified argument for the proof.
In this case, evaluating \PII\ at $z_0$ gives
a finite nonzero lower-bound for $|w''(z_0)|$.
Therefore, there exists an ascent path for $|w'|$ starting
at $z_0$. Following this path (over a bounded distance)
takes the solution to a point $z_1$ where $P(w)$
is lower-bounded by $|\epsilon|^{1/7}$. We can now start again
at such a point and
apply the theorem. Such arguments are used in more detail in section 3.
\item In the proof above, $B$ is not required to be finite, but it should
be {\em appropriate}. As remarked after Definition \ref{app},
$B$ may be quite large. For example, the error estimates of
the theorem are still small for $B\approx\epsilon^{-1/4}\approx\alpha^{1/8}$.
}
The last remark above shows that for any finite $z_0$ and large
but appropriate $B$, a generic solution $w$ has the {\em same}
elliptic-function-behaviour throughout the \lq\lq local\rq\rq\  domain $\Z$.
The error, however, does increase with $B$. If $B$ increases beyond
appropriate values, $z$ encroaches on the second asymptotic region
(i.e. $z\sim\alpha$) which will form the subject of a subsequent study.

\section{Limits of Generic Solutions}
In this section, we consider the possible asymptotic behaviours of
generic solutions $w(z)$ when the parameter $E$ or initial values
$\eta$, $\eta'$ become close to
degenerate values or become large.

Degenerate limits occur when
$E-D_k=:\delta^2$, with $|\delta|\ll 1$,
for some $k=1$, $2$ or $3$.
We separate this into the following
subcases.
\begin{enumerate}
\item For some integer $p>1$, $|\epsilon|^{p/7}<|\delta|\le |\epsilon|^{1/7}$
\item $|\delta|\le |\epsilon|^{p/7}$, for any integer $p>1$
\item For some $k=1$, $2$ or $3$,
$|\eta'|\le |\epsilon|^{1/7}$, $|\eta-d_k|\le |\epsilon|^{1/14}$
\end{enumerate}
Note that these cases overlap.
The first subcase is studied in Subsection 3.1 and the remaining
subcases are studied in Subsection 3.2 below.

The results of the direct proof of the \Pa\ property of \PII\
\cite{njmdk:direct} show that large initial values with finite $E$ imply that
the initial point $z_0$ is close to
a pole of the generic solution $w(z)$. The proof is a trivial rewriting
of the proof
given in section 3 of \cite{njmdk:direct} and so we do not reproduce it here.
In this case, Theorem \ref{genthm} shows that the pole of the \Pa\ transcendent
nearest to $z_0$ lies close to a pole of the
elliptic-function-type asymptotic behaviour of $w(z)$.

Therefore, the limit of interest, that violates the
genericity condition, is when $E$ becomes unbounded.
(Equivalently, we may take $\eta$, $\eta'$ large
with $\eta'\not=\eta^4$.)
In Subsection 3.3, we consider the large energy limit when
$|E|\ge |\ln\epsilon|^4$.
\subsection{Nearly Degenerate Solution}
The results of Theorem \ref{genthm} continue to apply even if
$\delta$ is smaller than $\epsilon^{1/7}$. (See Remark 1 following
Theorem \ref{genthm}.)
However, for
small enough $\delta$, the elliptic-function-type behaviours of
$w(z)$ are more accurately represented by singly periodic behaviours.
We show this here.

{}From the results of Section 2.1, we have that if $E=D_k+\delta^2$, where
$\delta\ll 1$,
\[
P(w)=(w-d_k)^2(w-\rho) (w-\sigma)+2\delta^2.
\]
Recall $d_k=(-1/2)^{1/3}$, and
$\rho$, $\sigma$ equals one of
\[
{1\over 2d_k^2}\left\{1\pm i\sqrt{2}\right\}.
\]

\begin{defn}
Let $\{\epsilon, z_0, B\}$ be an appropriate set, $k=1$, $2$, or $3$,
and $1<p$ be a given
number. Parameters
$\eta$,
$\eta'$ are called {\em nearly degenerate} if for
\[2E= \eta'^2 - \eta^4-2\eta,\quad E-D_k=:\delta^2,\]
they satisfy
\[
|\epsilon|^{1/7}<|\eta'|<|\ln\epsilon|^2,\quad
|\epsilon|^{1/14}<|\eta-d_k|<|\ln\epsilon|
\]
\[
|\epsilon|^{p/7}<|\delta| \le |\epsilon|^{1/7}
\]
\end{defn}
Note that the condition on $E$ implies $\eta'$, $\eta$ lie in a one-complex
dimensional deleted neighbourhood which excludes
the point $\eta'=0$, $\eta=d_k$, for a given $k$.
This deleted neighbourhood of initial conditions gives rise
to a solution that is
closer to a singly periodic limit of the previous elliptic function.
\begin{defn}
Suppose $\epsilon$, $z_0$ are appropriate numbers and $\eta$, $\eta'$ are
nearly degenerate parameters. Then the solution $w(z)$ of \PII\  given by
$w(z_0)=\eta$, $w'(z_0)=\eta'$ is called a {\em nearly degenerate} solution.
\end{defn}
Now the appropriate polynomial to consider is
\[R(w):=(w-d_k)^2(w-\rho) (w-\sigma),\]
and the $\mathcal W$-domain becomes
\[{\mathcal V} = \{ w\Bigm| |w|< |\ln\epsilon|, |R(w)|>\epsilon^{1/7}\}. \]
Correspondingly, $\epsilon$ is called small enough if the preimage
$\Z_\delta$ of $\mathcal V$ in $\mathcal Z$ is connected.
To be specific, we take $k=1$, with real
\[d_1=-1/2^{1/3}.\]

The main result in this case is
\begin{theorem}
Given an appropriate set $\{\epsilon, z_0, B\}$, with
$\kappa |\epsilon|^{5/7}|\ln\epsilon|^2\le 1$, where $\epsilon$ is small
enough, and nearly degenerate parameters
$\eta$,
$\eta'$, the nearly degenerate solution satisfies
\[
\left|w'^2 - R(w)\right|< 3|\epsilon|^{2/7},
\]
\[
\left|\int_\eta^w\,{dv\over(v-d_1)\sqrt{(v-\rho) (v-\sigma)}}
   -(z-z_0)\right| <6\pi B{|\epsilon|}^{1/7},
\]
for all $z\in\Z_\delta$.
\label{ndeg}
\end{theorem}
\noindent\textbf{Proof:} The proof is almost identical to that of
Theorem \ref{genthm} with a modified small term. Note that
\[w'^2 = R(w) + 2\delta^2 + 2S,\]
where $S$ is as defined in Eqn (\ref{s}). Therefore, since we still have
\[ 2|S|\le \kappa |\epsilon||\ln\epsilon|^2\]
whereas
\[ |\delta|^2 \le |\epsilon|^{2/7},\]
we get the first result. Integrating once more, we get
\beq
 \int_\eta^w\,{dv\over\sqrt{R(v)}}=\int_{z_0}^z\,dz
                  \left\{1+{2\delta^2\over R(w)}+{2 S\over
R(w)}\right\}^{1/2}.
\eeq
By assumption, we have
\[\left|{2\delta^2\over R(w)}\right|<2|\epsilon|^{1/7}\]
and
\[\left|{2 S\over R(w)}\right|<\kappa |\epsilon|^{6/7}|\ln\epsilon|^2.\]
So we have
\[\left|{2\delta^2\over R(w)}+{2 S\over R(w)}\right|<3
 |\epsilon|^{1/7}.\]
The remainder of the proof of Theorem \ref{genthm} can be repeated with
$\{2\delta^2+2 S\}/R(w)$ in place of $2 S/P(w)$ and $R(w)$ in
place of $P(w)$ to get the desired result.\quod

Note that we now require a smaller $B$ to get asymptotic validity.
Namely, we need $6\pi B{\epsilon}^{1/7}<<1$. For example,
$6\pi B<|\epsilon|^{-1/14}$ would be sufficient.

The interesting difference from the generic case arises because
the integral on the left can be evaluated explicitly in terms of
hyperbolic functions. To see this, make the change of variables:
\be
v&=&{(\sigma-\rho)\over 2}\biggl({f+f^{-1}\over 2}\biggr)+{(\sigma+\rho)\over
2}\\ f&=&\sqrt{3}g-\sqrt{2}i
\ee
Then the integral becomes
\[
\int_\eta^w\,{dv\over(v-d_1)\sqrt{(v-\rho) (v-\sigma)}}
    = \,{2\over d_1\sqrt{6}}i\int {dg\over g^2+1}.
\]
Hence we get
\beq
g = i\tanh\left\{{\sqrt{6}d_1\over 2}(z-z_0)+a_0+O({\epsilon}^{1/7})\right\},
\label{tanh}
\eeq
where $a_0=\tanh^{-1}\bigl(g(z_0)\bigr)$.
Note that this behaviour is singly periodic and possess a line (rather than
a lattice) of poles. In a direction orthogonal to the line, the behaviour
becomes asymptotic to constants.

There is an alternative change of variables.
Consider the transformation:
\be
v&=&d_1+u\\
u&=&-1/F\\
F&=&{1\over\sqrt{18d_1^2}}G +\,{1\over 3d_1}
\ee
Then the integral becomes
\[
\int_{\eta-d_1}^{w-d_1}\,{du\over u\sqrt{u^2+4d_1u+6d_1^2}}
={1\over\sqrt{6}d_1}\int {dG\over\sqrt{G^2+1}}.
\]
In this case, we get
\beq
G=\sinh\left\{\sqrt{6}d_1(z-z_0)+b_0+O({\epsilon}^{1/7})\right\},
\label{sinh}
\eeq
where $\sinh b_0=G(z_0)=-\sqrt 2 +\sqrt{18}d_1/(d_1-\eta)$.
The behaviours indicated by Eqns(\ref{tanh}) and (\ref{sinh}) are
clearly related and represent the same nearly degenerate solution
in the domain $\Z_\delta$.
Note that one line of poles is given approximately by
\beq z_0+{-b_0+2\pi n i\over\sqrt 6 d_1}\label{chain}\eeq
for integer $n$.

In Appendix A, we show that one of the periods $\om_j$ of the
generic elliptic-function behaviour becomes logarithmically large
and one remains finite as $E\to D_k$ for some $k=1, 2, 3$.
For the nearly degenerate case this shows that for some $j$
\[
\om_j \sim \ln |E-D_k|\sim |\ln\epsilon|.
\]
For sufficiently small $\epsilon$,
this is clearly smaller than $B$ (e.g. $B\sim |\epsilon|^{-1/14}$)
for which the asymptotic results
of Theorem \ref{ndeg} are valid.
Therefore, the results above suggest
that more than one line of poles may be visible in the domain $\Z_\delta$.
Between the lines of poles, the local behaviour is still described by
Eqn (\ref{sinh}).

\subsection{Degenerate Solution}
Here we consider the case when $\delta$ is arbitrarily
small or when initial values
are arbitrarily close to degenerate values.
\begin{defn}
Let $\{\epsilon, z_0, B\}$ be an appropriate set, and, for
some $k= 1, 2, 3$, let
$\eta$, $\eta'$ be given parameters.
\bi\item If they satisfy
\[2E:= \eta'^2 - \eta^4-2\eta =2D_k+2\delta^2,\]
where $|\delta|\le |\epsilon|^{p/7}$ for all integers $p>1$, and
\[
|\epsilon|^{1/7}<|\eta'|<|\ln\epsilon|^2,\quad
|\epsilon|^{1/14}<|\eta -d_k|<|\ln\epsilon|
\]
then they are called {\em almost degenerate} parameters.
\item If, on the other hand,
they satisfy
\[|\eta'|\le |\epsilon|^{1/7}, \quad |\eta-d_k|\le |\epsilon|^{1/14}\]
then they are called {\em degenerate} parameters.
\ei
\end{defn}
To be concrete, we take $k=1$. Assume that
$\epsilon$, $\mathcal V$, $\Z_\delta$ are as defined in the
previous section, with a slight modification that $|R|\ge |\epsilon|^{1/7}/16$
on $\Z_\delta$.
\begin{defn}
Let $\epsilon$ be {\em small enough} such that the domain $\Z_\delta$
is connected and let $\eta$ $\eta'$ be almost degenerate parameters.
Then the solution $w(z)$ of \PII\  given by
$w(z_0)=\eta$, $w'(z_0)=\eta'$ is called an {\em almost degenerate} solution.
\end{defn}
The solution defined by almost degenerate parameters has asymptotic behaviour
that is given locally by a singly periodic function.
In fact, the proof of the last subsection can be applied to show that
the following theorem holds.
\begin{theorem}
Given an appropriate set $\{\epsilon, z_0, B\}$, with
$\kappa |\epsilon|^{5/7}|\ln\epsilon|^2\le 1$, where $\epsilon$ is small
enough, and almost degenerate parameters
$\eta$,
$\eta'$, the almost degenerate solution satisfies
\[
\left|w'^2 - R(w)\right|< {\kappa\over 2}|\epsilon|\,|\ln\epsilon|^2
\]
\[
\left|\int_\eta^w\,{dv\over(v-d_1)\sqrt{(v-\rho) (v-\sigma)}}
   -(z-z_0)\right| <2\pi \kappa B
|\epsilon|^{6/7}|\ln\epsilon|^2.
\]
for all $z\in\Z_\delta$.
\label{adeg}
\end{theorem}
Note that the domain of validity is now larger than in Theorem \ref{ndeg},
i.e. $B$ may be as large as $1/|\epsilon|^{5/7}$.

The almost degenerate case clearly also has representations given by
Eqns(\ref{tanh}) and (\ref{sinh}) in the domain
$\Z_\delta$. However, now there may be at most one line of poles.
The reason lies in the fact that the almost degenerate limit
of an elliptic-function behaviour has a period that is
possibly $\gg |\ln\epsilon|$.
For example, suppose $\delta\sim\exp(-1/|\epsilon|)$. Then
\[
\om_j\sim \ln(E-D_k)\sim 1/|\epsilon|.
\]
Hence the distance between two lines of poles (or any two lines
formed by interpolation through integer
multiples of the finite period) may be much larger than
appropriate values of $B$.

Nevertheless, there must be at
least one line of poles in the local domain,
because the lower bound on $\eta-d_1$
implies $|b_0|<|\ln\epsilon|$. Hence the sequence given by
Eqn(\ref{chain}) lies in $\Z_\delta$.

The remaining case, apparently different to this, is the degenerate
case. However, we can show that the almost degenerate solution
becomes a degenerate solution somewhere in its domain of
validity.
To show this, we travel on a descent path.
Recall that a complex-valued
function $f(z)$ of one complex variable $z$ possesses a descent path
(for its modulus) starting a point $z_0$ if $f'(z_0)\not=0$.
Along the steepest descent path, we have $|df|=-d|f|$.

\begin{theorem}
Let $\{\epsilon, z_0, B\}$ be an appropriate set with
$|\epsilon|^{1/4}<2^{-1/3}$.
Let $\eta$, $\eta'$ be almost degenerate parameters and
let $w$ be an almost degenerate solution. Then
there exists a path $\Gamma$ joining $z_0$ to a point $z_1$ where
$|w-d_k|^2$ and $|w'|$ become less than
$|\epsilon|^{1/7}$, for some $k=1, 2, 3$.
\end{theorem}
\nbf{Proof:}
{\em Case 1:} Consider the case when $|\eta|\ge |d_k|/4$.
By assumption of appropriateness and upperbounds on $\eta$, $\eta'$, we have
\[\left|{\eta+z_0\eta'\over\alpha}\right|< 2|\epsilon||\ln\epsilon|^2.\]
Therefore, we can lower-bound
\beq w''' = 6w^2w' + {1\over\alpha}\bigl(w+z w'\bigr)\label{3}\eeq
at (and by continuity near) $z_0$ by
\[|w'''(z_0)| > |d_k|^2|\epsilon|^{1/7}/4.\]
In other words, there exists a descent path for $w''$ starting
at $z_0$. We take $\Gamma$ to be this descent path.

Recall that in \PII\
\beq w'' = 2w^3 + 1 +{z\over\alpha}w\label{2}\eeq
and its first integral
\beq
w'^2 = R(w) + 2S, \label{1}
\eeq
we have
\beq
\left|{z\over\alpha}w\right| < 3|\epsilon|\,|\ln\epsilon|,\quad
|S| < \kappa/2 |\epsilon|\,|\ln\epsilon|^2\label{small}
\eeq
throughout $\Z_\delta$, and in particular on the part of the
path $\Gamma$ lying in $\Z_\delta$.

Since $w''$ is decreasing, this shows that
$2w^3+1$ must decrease along $\Gamma$, until it reaches
a first point $z_1$ where $|w-d_k|$ becomes
smaller than $|\epsilon|^{1/14}$, or where we reach the
boundary of $\Z_\delta$. Similarly $R(w)$ also
decreases along $\Gamma$.

At that point, the definitions of
$d_k$, $\rho$, and $\sigma$ show that
\[\Bigl|R\bigl(w(z_1)\bigr)\Bigr| \ge {|\epsilon|^{1/7}\over 16},\]
while it remains larger than this bound earlier on $\Gamma$.
Moreover, we can estimate its length.
For conciseness,
we rename $v:=w''$. By definition
we get
\bes
z_1 - z_0 &=& \int^{z_1}_{z_0} dz\\
          &=& \int_{\Gamma}\,{dv\over v'}
\ees
But from Eqns (\ref{3}), (\ref{1}), and (\ref{small}), we get
\[|v'|\ge 4|w|^2|w'| \ge |d_k|^2|\epsilon|^{1/14}/64\]
if
\beq
|\epsilon|^{1/14}> 32\pi^2 d^2 |\epsilon|\,|\ln\epsilon|
\label{ub}
\eeq
where $d$ is the
length of $\Gamma$. We then get
\bes
d &=& \Bigl|\int_{\Gamma}\,{dv\over v'}\Bigr|\\
  &\le& \int_{\Gamma}\,{|dv|\over |v'|}\\
  &\le& - {64\over |d_k|^2}\int_{\Gamma}\,d|v|/|\epsilon|^{1/14}\\
  &=&{64|\epsilon|^{-1/14}\over |d_k|^2} \Bigl(\bigl|v(z_0)\bigr| -
\bigl|v(z_1)\bigr|\Bigr)
\ees
This clearly satisfies Eqn(\ref{ub}). Moreover, this shows that
$\Gamma$ lies within $\Z_\delta$.

\noindent{\em Case 2:} Now consider the case when $|\eta|<
|d_k/4|$.
We have from
\[ w'' = 2w^3 + 1 +{z\over\alpha}w\]
that
\[|w''(z_0)| \ge 1/2.\]
Hence there exists a descent path for $w'$ starting at $z_0$.
We take $\Gamma$ to be this path.
The decrease in $|w'|$ implies a corresponding decrease in $|R(w)|$.
Therefore, $w$ must get closer to a root of $R(w)$. Since all such roots
have modulus $\ge 1/2^{1/3}$,
$|w|$ must increase.
There exists a first point $\tilde z_0$ on $\Gamma$ where $w$
has modulus equal to $|d_k/4|$.
That is,
\[\left|w(\tilde z_0)-d_k\right| \ge {3|d_k|\over 4}.\]
At such a point, we have
\[ |R(w)| > 18 |d_k|^2,\]
by using the definitions of $d_k$, $\rho$, and $\sigma$.
Using arguments similar to above we get
\[|\tilde z_0 -z_0|=:d \le {1\over2\sqrt{18}}.\]
We now use the results of Case 1 starting at $\tilde z_0$ to
extend $\Gamma$ to $z_1$.
\quod

It is instructive to consider the limits of Theorems \ref{ndeg}, \ref{adeg},
explicitly when
\[0\not= |w-d_k| < |\epsilon|^{1/14}, \quad
0\not=|\eta-d_k| < |\epsilon|^{1/14}.\]
We have
\bes
& &\int_\eta^w\,{dv\over\sqrt{R(w)}}\\
 \quad&=&{1\over\sqrt 6d_k}\int_\eta^w\,{dv\over (w-d_k)\sqrt{1+2(w-d_k)/(3d_k)
                                           +(w-d_k)^2/(6d_k^2)}}\\
 \quad&=&{1\over\sqrt 6d_k}\int_\eta^w\,{dv\over (w-d_k)}\Biggl\{1 -
(w-d_k)/(3d_k)
                                          +O\bigl((w-d_k)^2\bigr)\Biggr\}\\
 \quad&=&{1\over\sqrt 6d_k}\Biggl\{\ln\Bigl((w-d_k)/(\eta-d_k)\Bigr)
                              - (w-\eta)/(3d_k)
                              + O\bigl((w-d_k)^2\bigr)\Biggr\}
\ees
Taking the inverse of the second result of Theorem \ref{adeg} implies
that
\be
w-d_k &=& (\eta-d_k)\exp\left(\sqrt 6d_k(z-z_0)\right)
         \left(1 - (\eta-d_k)/(3d_k) + O(|\epsilon|^{1/7})\right)\nonumber\\
	\quad&+& {1\over 3d_k}(\eta-d_k)^2\exp\left(2\sqrt 6d_k(z-z_0)\right)
        \bigl(1 + O(|\epsilon|^{1/14})\bigr).\label{osc}
\ee
It is clear that the exponential terms here will grow in a half-plane
of directions in $\Z_\delta$.
For example, consider $d_1 = -1/2^{1/3}$. Then
\[\Re(z-z_0)<0\ \Rightarrow\ \Re\bigl(d_1(z-z_0)\bigr)>0.\]
In this half-plane, the solution grows away from the initial assumed value.
By following an ascent path, it is straightforward to show that this
behaviour matches with that given by Eqn(\ref{sinh}) in this
half-plane.

Conversely, if the solution has degenerate initial values, we can
integrate \PII\ to show that $w$ is given by Eqn(\ref{osc}) in a local
domain around $z_0$. To prove this, we consider a domain given by
\[
{\mathcal U} = \{ w\Bigm| |\epsilon|^{1/7}<|w|< |\ln\epsilon|,
|w-d_k|<\epsilon^{1/14}\}.
\]
Correspondingly, define $\epsilon$ to be small enough if the preimage
$\Z_0$ of $\mathcal U$ in $\mathcal Z$ is connected.
\begin{defn}
Let $\epsilon$ be {\em small enough}
and let $\eta$, $\eta'$ be degenerate parameters.
Then the solution $w(z)$ of \PII\  given by
$w(z_0)=\eta$, $w'(z_0)=\eta'$ is called a {\em degenerate} solution.
\end{defn}
Then we have the following theorem.
\begin{theorem}
Assume we are given an appropriate set $\{\epsilon, z_0, B\}$, where $\epsilon$
is small
enough such that
\[
\kappa |\epsilon|^{5/7}|\ln\epsilon|^2\le 1,\
|\epsilon|^{1/7}<|d_k|,
\]
and
\[
6|d_k|^2+4|d_k||\epsilon|^{1/14}+|\epsilon|^{1/7}< 7|d_k^2|.
\]
Furthermore, assume that degenerate parameters
$\eta\not=d_k$,
$\eta'\not=0$, are given. Then there exists a number $\mu>0$ such that
the degenerate solution satisfies
\[\left|w'^2 - 6d_k^2(w-d_k)^2\right|\le 14|d_k|^2|\epsilon|^{1/7},\]
\[
\left|\int_\eta^w\,\,{dv\over\sqrt{6d_k^2(v-d_k)^2+2\delta^2}}
   -(z-z_0)\right| < 2\pi B\mu|\epsilon|^{1/14}.
\]
for all $z\in{\Z_0}$.
\end{theorem}
\nbf{Proof:} Again, the technique of the proof is almost the
same as Theorem \ref{genthm} and we only highlight the important differences.
After one integration, we get
\[ w'^2 = R(w) + 2\delta^2 + 2S,\]
where by the given assumptions, we have
\[ 2|\delta|^2= 2|E-D_k| \le 8|d_k|^2|\epsilon|^{1/7}.\]
$S$ here is as before, with the same upperbound:
$|2S|<\kappa |\epsilon||\ln\epsilon|^2$.
To get the first result, we use the fact that
\[ R(w) = 6d_k^2(w-d_k)^2 + 4d_k(w-d_k)^3+(w-d_k)^4,\]
and the bounds on $w-d_k$.

Hence it is slightly more transparent to write
\[ w'^2 = 6d_k^2(w-d_k)^2 + 4d_k(w-d_k)^3+(w-d_k)^4+2\delta^2 + 2S.\]
The second result follows from taking the square root, dividing by
\hspace\fill\break
$\sqrt{6d_k^2(w-d_k)^2+2\delta^2}$ and bounding the remaining error terms.
\quod
\remarks{
\item Note that in the case $\eta=d_k$, $\eta'=0$, $E=D_k$ and the
integral in the second result above is divergent. We consider this
case here.
At $z_0$, we have
\[ w''(z_0) = z_0\eta/\alpha, w'''(z_0) = \eta/\alpha,\]
and so there exists an ascent path for $w''$ starting at $z_0$.
By \PII , this is also an ascent path for $2w^3+1$. Following this
path, even for a short distance (say of order $\epsilon^{1/14}$) allows
us to reach a point $z_1$, say, where we can
use the above argument for small but nonzero $w(z_1)-d_k$.
\item The above theorem extends easily to the case
when $\eta\not=d_k$ but $\eta'=0$, because $E-D_k\not=0$.
\item
The case when $\eta=d_k$, but $\eta'\not=0$ can be treated by an
argument similar to the above ascent-path argument.
\item
The only remaining difficulty is that the path of integration may originate at
or possibly intersect one of the simple zeroes of $R(w)$.
(Note that the definition
of $\mathcal U$ does not exclude the simple zeroes $\rho$,
$\sigma$ of $R(w)$ from the path.) However, these problems are clearly
overcome by ascent-path-type argument similarly to the way in which the
zeroes of $P(w)$ were avoided in Remark 4 following the
proof of Theorem \ref{genthm}.}

These remarks cover all the remaining degenerate possibilities.
\subsection{Large Energy Limit}
Here we consider the only remaining case of possible limits of the generic
solution, that is, we consider the case when $|E|$ (or equivalently
$\eta$ or $\eta'$, with $\eta'\not=\eta^4$) is large.
\begin{defn}
Suppose $\epsilon$ is an appropriate number. Parameters $\eta$, $\eta'$
are called {\em large} if
\[2E = \eta'^2 -\eta^4 - 2\eta\]
satisfies
\[ |E|\ge |\ln\epsilon|^4.\]
The corresponding solution $w$ defined by initial values $w(z_0)=\eta$
$w'(z_0)=\eta'$ is called a {\em large} solution.
\end{defn}
In this case, it is convenient to rescale variables as follows. Let
\bes
w &:=& E^{1/4} q(s)\\
s &:=& E^{1/4} z
\ees
Then the first integral of \PII\ becomes
\beq
q_s^2 = q^4+2 +{2\over E^{3/4}}\Bigl(q + \int ds\,s\,qq_s/\alpha\Bigr).
\label{firstq}
\eeq
\begin{defn}
Define the $q$-domain
\[{\mathcal Q} :=\Bigl\{ sE^{-1/4}\in{\mathcal Z}\Bigm||q|<|E|^{1/4},
         |q^4+2|>|E|^{-1/4}\Bigr\}\]
and $s$-domain
\[
\Sd := \{ sE^{-1/4}\in{\mathcal Z}\Bigm| q(s)\in \mathcal Q\}.
\]
$\epsilon$ is called {\em small enough} if $\Sd$ is connected.
\end{defn}
Let $s_0=z_0E^{1/4}$, $\iota = \eta E^{-1/4}$, $\iota' = \eta'E^{-1/2}$.
We will loosely call $\iota$, $\iota'$, and
$q(s)=E^{-1/4}w(sE^{-1/4})$ {\em large}
if $\eta$, $\eta'$ and $w(z)$ are large.

Note that we will now restrict paths of integration to those with
length upperbounded by $2\pi B$ in $\Sd$.
\begin{theorem}
Given an appropriate set $\{\epsilon, z_0, B\}$, with
$|\epsilon\kappa|/2 < 1/2$ where
$\epsilon$ is small enough,
and large $\iota$, $\iota'$, and the corresponding solution $q(s)$
satisfies
\bes
\left|q_s^2 -(q^4+2)\right|&\le& {3\over 2}|E|^{-1/2}\\
\left|\int_\iota^q\,{dq\over\sqrt{q^4+2}
   }-(s-s_0)\right| &\le& 3\pi B |E|^{-1/2}
\ees
for all $s\in\Sd$. Moreover, if $s_0$ and $s_{10}$ are two
successive points in $\Sd$ where $w=\iota$, then for
$j=1$ or $2$,
\[
\left|(s_{10}-s_0)-\omega_j\right|\le 3\pi B |E|^{-1/2}.
\]
\end{theorem}
\nbf{Proof:}
The small term in the first integral Eqn(\ref{firstq})
is given by
\[T:= {1\over E^{3/4}}\Bigl(q + sq^2/(2\alpha)-\int ds\,q^2/(2\alpha)\Bigr).\]
The results follow from the techniques of the proof of Theorem \ref{genthm}.
\quod

\section[]{Acknowledgements}
It is a pleasure to thank the
Isaac Newton Institute, where this study was first started, and the Australian
Research Council for their support.
\appendix\section{Complete Elliptic Integrals}
Here we prove some results for complete elliptic integrals as functions
of $E$.

Let $C$
be one of $C_k$, $k= 1, 2$.
\begin{lemma}
The elliptic integrals
\[\tilpsi=\oint_{C}dw\,{w^2\,\sqrt{w^4+2w+2E}},\]
\[\psi=\oint_{C}dw\,{w^2\over\sqrt{w^4+2w+2E}},\]
\[\psi'=-\oint_{C}dw\,{w^2\over{w^4+2w+2E}^{3/2}},\]
considered as functions of $E$ satisfy the following differential equations.
\be
{d\tilpsi\over dE}&=&\psi\nonumber\\
{d\psi\over dE}&=&\psi'\nonumber\\
{d^2\tilpsi\over dE^2}&=&\biggl({40E\over 128E^3-27}\biggr)\tilpsi
\label{pside}
\ee
\end{lemma}
\noindent\textbf{Proof:} The first two statements are obvious. To prove the
last assertion we use the following definitions of other elliptic integrals.
\bes
\tildo&:=&\oint_{C}dw\,{\sqrt{w^4+2w+2E}}\\
\om&:=&\oint_{C}dw\,{1\over\sqrt{w^4+2w+2E}}\\
\primo&:=&-\oint_{C}dw\,{1\over{w^4+2w+2E}^{3/2}}\\
\tilphi&:=&\oint_{C}dw\,{w\,\sqrt{w^4+2w+2E}}\\
\phi&:=&\oint_{C}dw\,{w\over\sqrt{w^4+2w+2E}}\\
\phi'&:=&-\oint_{C}dw\,{w\over{w^4+2w+2E}^{3/2}}
\ees
That these are interrelated can be shown by integration by parts as follows:
\bes
\tilpsi&=&-{1\over 3}\oint_Cdw\,w^3\, {2w^3+1\over\sqrt{P(w)}}\\
       &=&-{2\over 3}\oint_Cdw\,w^2\, {w^4+2w+2E-(3/2)w-2E\over\sqrt{P(w)}}\\
       &=&-{2\over 3}\tilpsi+\oint_Cdw\,
{w^3\over\sqrt{P(w)}}+{4E\over 3}\psi\\
       &=&-{2\over 3}\tilpsi+{1\over 2}\oint_Cdw\,
{2w^3+1\over\sqrt{P(w)}}-{1\over 2}\om+{4E\over 3}\psi
\ees
Therefore, we get
\beq \om={8\over 3}E\psi-{10\over 3}\tilpsi.
\label{A1}
\eeq
Similarly, starting with $\tildo$ and integrating by parts gives
\beq
3\tildo=3\phi+4E\om ,
\label{A2}
\eeq
and starting with $\tilphi$ gives
\beq
2\tilphi={3\over 2}\psi+2E\phi .
\label{A3}
\eeq
Finally, starting with $\phi$, we get
\beq
\phi'=-\,{4\over 3}E\phi '.
\label{A4}
\eeq
Differentiating Eqn(\ref{A1}) w.r.t. $E$ yields
\beq
\primo={8\over 3}E\psi'-{2\over 3}\psi.
\label{A1p}
\eeq
Doing the same with Eqn(\ref{A2}) gives
\beq
-\om=3\phi'+4E\primo.
\label{A2p}
\eeq
Now eliminating $\tildo$, $\om$, $\primo$ between Eqns(\ref{A1}-\ref{A2p})
gives the desired differential equation for $\psi(E)$. \quod

Note that these show that $\tildo_j$ and, therefore, the periods $\om_j$
are singular at the zeroes of $128E^3-27$. These are precisely the
degenerate values $D_k$ found in Eqn(\ref{Dk}).

Consider $\tilpsi_j$, in the degenerate limit $E\to D_k$ (for some $k$).
We drop the subscript $j$ below, and consider momemtarily
only the singular case.
Fr\"obenius expansion of the solutions $\tilpsi$ of Eqn(\ref{pside})
show that there exists a singular solution with behaviour
\[ \tilpsi \approx c_0 + c_1 (E-D_k)\ln(E-D_k)\]
as $E\to D_k$, where $c_0$ is an arbitrary constant and $c_1$ is a multiple
of it. (For a derivation of similar results with more detail see
Appendix B of \cite{njmdk:conn1}.)
Moreover, we have
\[\psi = \tilpsi'(E) \sim c_1\ln(E-D_k)\]
as $E\to D_k$.
The period in this case is given by
\beq
\om={8\over 3}E\psi-{10\over 3}\tilpsi \approx {8D_kc_1\over 3} \ln(E-D_k).
\label{logomega}
\eeq

There is also a linearly dependent solution analytic at $D_k$ with expansion
\[\om_j\approx b_0+b_1 (E-D_k) \]
as $E\to D_k$, where $b_0$ is an arbitrary constant and $b_1$ is a
multiple of it.

\end{document}